\documentclass[12pt]{article}

\newcommand{\bibl}[5]
	{#1, {\it #2} {{\bf #3},} #5 (#4)}
\newcommand{\cebe}{\begin{center}}
\newcommand{\ceen}{\end{center}}
\newcommand{\debe}{\begin{description} \vspace{-2ex}}
\newcommand{\deen}{\end{description}}
\newcommand{\eabe}{\begin{eqnarray}}
\newcommand{\eaen}{\end{eqnarray}}

\newcommand{\eqbe}{\begin{equation}}
\newcommand{\eqen}{\end{equation}}

\newcommand{\itbe}{\begin{itemize}}
\newcommand{\iten}{\end{itemize}}

\newcommand{\tabe}{\begin{tabbing}}
\newcommand{\taen}{\end{tabbing}}

\usepackage{psfig}

\begin{document}

\begin{titlepage}
\begin{flushright}
  LU TP 98-14 \\
  hep-ph/9807541 \\
  July 1998
\end{flushright}
\vspace{25mm}
\begin{center}
  \Large
  {\bf Is there screwiness at the end \\ of the QCD cascades?} \\
  \normalsize
  \vspace{12mm}
  Bo Andersson, G\"osta Gustafson, Jari H\"akkinen, \\ Markus Ringn\'{e}r and
  Peter Sutton\footnote{bo@thep.lu.se, gosta@thep.lu.se, jari@thep.lu.se, 
	markus@thep.lu.se, peter@thep.lu.se}
  \vspace{1ex} \\
  Department of Theoretical Physics, Lund University, \\
  S\"olvegatan 14A, S-223 62 Lund, Sweden \\
\end{center}
\vspace{20mm}

\noindent {\bf } \\ 
 We discuss what happens at the end of the QCD cascades.  We show
that, with just a few reasonable assumptions, the emission of soft
gluons is constrained to produce an ordered field in the form of a
helix.  We describe how to modify the Lund fragmentation scheme in
order to fragment such a field. Our modified fragmentation scheme
yields results which are consistent with current experimental
measurements, but predicts at least one signature which should be
observable.

\vspace{3cm}

\end{titlepage}

\section{Introduction} \vspace{-2ex}
In QCD the production of two colour charges which subsequently move
apart will lead to the production of further colour radiation. This
can be described in terms of the fundamental field quanta, the gluons,
but it is also possible to describe the ensuing radiation in terms of
dipoles. This property arises because in non-abelian theories the
emission of an extra gluon from a gluon-gluon dipole can (to a very
good approximation) be modelled as the destruction of the original
dipole and the creation of two new dipoles. In this way the change in
the colour field can be described as an increasing cascade of dipoles.
The end of this cascade occurs when the dipole masses are so small
that helicity conservation prevents further real gluon emission. In
this paper we examine what happens at the end of this cascade. We
find that the conditions are favourable for the field to utilize the
azimuthal degree of freedom and wind itself into the form of a helix.
This corresponds to a close--packed configuration of gluons in
rapidity--azimuthal-angle space.

We begin by describing a toy model which contains the relevant
features, namely a tendency to emit as many gluons as possible and the
constraint that gluons are not too ``close'' to each other (which
arises from helicity conservation).  In this simple model it is clear
that at the end of the cascade an ordered field emerges with the
characteristics of a helix.
To progress beyond this model we use the Lund model of QCD.
In the Lund picture hard gluons are represented as excitations of a
relativistic string which connects a quark, anti-quark pair. However,
the gluons from which the helix is built up are too soft to be
modelled in this way. Instead we introduce a helical semi-classical
field and thus develop a modifed version of the Lund fragmentation
scheme.  Our modified fragmentation scheme enables us to study
whether the consequences of a screwy field can be detected in the final
state particles. We find that if events with hard gluons are
excluded then the screwiness of the field may be observed.

\section{The dipole cascades; increase and decrease of phase space}
\label{s:dipole}

In order to describe what can happen at the end of the QCD cascades we
will provide a brief description of the cascades. We will in
particular discuss the consequences of helicity conservation in the
emission of partons.

The well-known formula for dipole emission of bremsstrahlung is
\begin{eqnarray}
\label{e:bremsstrahlung}
dn= \bar{\alpha}\frac{dk_{\perp}^2}{k_{\perp}^2} dy (\frac{d\phi}{2\pi})\,\Psi
\end{eqnarray}
where $\bar{\alpha}$ is the effective coupling, $k_{\perp}$, $y$, and
$\phi$ are the transverse momentum, rapidity and azimuthal angle respectively,
although the azimuthal angle dependence is usually neglected. 
The final factor, $\Psi$, corresponds to the spin
couplings. We will briefly consider the precise definitions before we
consider the implications. The effective coupling for QCD in the
case of a gluonic dipole is given by
\begin{equation}
\label{e:couplings}
\bar{\alpha}_{\mbox{\scriptsize QCD}}  =  \frac{N_c \alpha_s}{2 \pi} \simeq \frac{6}{11
\log(k_{\perp}^2/\Lambda^2)}.
\end{equation}
The occurrence of the number of colours, $N_c$, and the factor $1/2$ in the
QCD coupling is due to early conventions, whereas the result that
the running is governed by $1/c = 6/11$ is a basic gauge group
independent result. It only depends upon the fact that in non-abelian
gauge theories there is a three-particle coupling between vector
particles, e.g. the colour-$8$ gluons in QCD. (The four-gluon
coupling also occurs to preserve the symmetry, but it does not play a
r\^{o}le in this connection).  We 
neglect the flavour term $-2n_f/3$ which should
accompany $11$ in the denominator because it is a small effect related
to the possibility of gluon splitting; $\mbox{g}\rightarrow
\mbox{q}\overline{\mbox{q}}$.

The transverse momentum and the (dipole cms) rapidity are
defined in a Lorentz invariant way in terms of the squared masses of
the final state partons (the emitters are conventionally indexed $1$
and $3$ and the emitted field quantum $2$):
\begin {eqnarray}
\label{e:kinvar}
s_{ij} & = & (k_i+k_j)^2 =2k_ik_j  =  2 k_{\perp i}k_{\perp j} \left[\cosh (\Delta
y)_{ij}-\cos(\Delta \phi)_{ij}\right] \nonumber \\
s & = & s_{12}+s_{23}+s_{31} \nonumber \\ 
s_{12} & = & s(1-x_3),~ s_{23}=s(1-x_1)  \nonumber \\
k_{\perp}^2 & = & \frac{s_{12}s_{23}}{s} \nonumber \\
y & = & \frac{1}{2} \log \left(\frac{s_{12}}{s_{23}}\right)
\end{eqnarray}
Here $x_1$ and $x_3$ are the final state cms energy fractions of the
emitters. Requiring energy momentum conservation limits the allowed
emittance region to
\begin{eqnarray}
\label{e:energycons}
k_{\perp}\cosh(y) \leq \frac{\sqrt{s}}{2}.
\end{eqnarray}
This region can conveniently be approximated as $|y| < (L
-\kappa)/2$ with the variables $L\equiv\log(s/\Lambda^2)$ and
$\kappa\equiv\log(k_{\perp}^2/\Lambda^2)$.
This means that the (approximate) phase space available for dipole
emission is the interior of a triangular region in the
$(y,\kappa)$-plane with the height and the baselength both equal to
$L$.  The inclusive density inside the triangle is, in this Leading-Log
Approximation~(LLA), given by the effective coupling $\bar{\alpha}$ according
to Eq.(\ref{e:bremsstrahlung}). The rapidity range, $L-\kappa$, is of
course the length of a hyperbola spanned between the emitters in
space-time (or energy-momentum at the scale $k_{\perp}^2$).

If we consider an initial $\mbox{q}\overline{\mbox{q}}$ dipole
emitting a gluon then the probability for the produced
$\mbox{qg}\overline{\mbox{q}}$ system to emit a second gluon is a
complicated expression~\cite{r:twogluon}.  In case the transverse
momenta of the first and second gluon are strongly ordered, $k_{\perp
1} \gg k_{\perp 2}$, it is a very good approximation to treat the
second emission as independent emission from two
dipoles~\cite{r:dipole}. For an exclusive statement, for example the
probability $dP$ to emit the first gluon with a certain
$(\kappa_1,y_1)$, it is necessary to multiply the inclusive formula in
Eq.(\ref{e:bremsstrahlung}) with a Sudakov form factor $\Delta_s$
containing the probability not to emit above $\kappa_1$,
\begin{eqnarray}
\label{e:Sudakov}
\Delta_s (L,\kappa_1) & = & \exp(-\int^L_{\kappa_1} dn), \nonumber\\
dP(\mbox{q},\mbox{g}_1,\overline{\mbox{q}}) & = & dn(\kappa_1,y_1) \Delta_s(L,\kappa_1).
\end{eqnarray}
The probability to emit two gluons is then, in the approximation that
the second gluon is emitted by two independent dipoles, given by
\begin{equation}
\label{e:2gluon}
dP(\mbox{q},\mbox{g}_1\mbox{g}_2,\overline{\mbox{q}})= dP(\mbox{q},\mbox{g}_1,\overline{\mbox{q}})\left[dP(\mbox{q},\mbox{g}_2,\mbox{g}_1)+dP(\mbox{g}_1,\mbox{g}_2,\overline{\mbox{q}})\right] 
\end{equation}
in easily understood notations.  The approximation in
Eq.(\ref{e:2gluon}) results at most in a percentage error over
all phase space~\cite{r:percentage}. Thus, contrary to QED where the
chargeless photons still leave the $\mbox{e}^+\mbox{e}^-$-current as
the single emitter, the $8$-charge gluon ($\mbox{g}_1$) in QCD changes
the original $\mbox{q}\overline{\mbox{q}}$ dipole into two dipole
emitters, one between $\mbox{q}$ and $\mbox{g}_1$ and one between 
$\mbox{g}_1$ and $\overline{\mbox{q}}$, and {\em each can independently emit
the second gluon }($\mbox{g}_2$).  The requirement for the validity of
the approximation in Eq.(\ref{e:2gluon}) is that $k_{\perp 1}\geq
k_{\perp 2}$ or else the indices are exchanged.

The two independent dipoles are moving apart (with $\mbox{g}_1$ as the
common parton). This means that they have together a larger effective
rapidity range for the emission of $\mbox{g}_2$, i.e. the original
hyperbola length $L=\log(s)$ is exchanged for two hyperbolas with the
combined length $\log(s_{\mbox{\scriptsize qg}_1})
+\log(s_{\mbox{\scriptsize g}_1\overline{\mbox{\scriptsize q}}})= L
+\log(k_{\perp 1}^2)$.  From any one of the two new dipoles we may then
emit the second gluon, thereby producing three independent dipole
emitters and the process can be continued towards more dipoles;
ordering the process in $k_\perp$ downwards. The available phase space
for further emission is increased after each emission, as can be seen
from the increased total length, $L$, after the first emission. This
description of the QCD cascades is called the Dipole Cascade Model
(DCM)~\cite{r:DCM}.

We will now consider the polarisation sum contribution in 
Eq.(\ref{e:bremsstrahlung}). Its precise properties depend upon whether
we are dealing with a $\mbox{q}\overline{\mbox q}$, $\mbox{qg}$
or a $\mbox{gg}$ dipole, but it stems
from the spin couplings between the emitter(s) (it is essentially
sensitive only to the closest emitter)
and the new field quantum. These couplings contain the property
that helicity is conserved, which is true for all gauge theories.
This means that if a spin-$1/2$ parton emits a spin-$1$ parton, the
spin-$1$ parton must go apart from the emitting particle in order to
conserve helicity and angular momentum. They have to go even further
apart in the case of a spin-$1$ parton emitting a spin-$1$ parton. To
estimate the separation we consider (for fixed $k_{\perp}$ (or
$\kappa$)) the available rapidity range:
\begin{eqnarray}
\label{e:MLLArap}
\int^{y_{\mbox{\tiny max}}}_{y_{\mbox{\tiny min}}}\Psi\, dy =L-\kappa - c +O(k_{\perp}^2/s)
\end{eqnarray}
where $c=(11/12 +11/12)$, $(3/4+11/12)$ or $(3/4+3/4)$ depending on
whether the emitters are $\mbox{gg}$, $\mbox{qg}$
($\mbox{g}\overline{\mbox{q}}$) or a $\mbox{q}\overline{\mbox{q}}$
dipole~\cite{r:gostamult}. The quantities $c$ are written as sums to
show that a spin-$1$ ($\mbox{g}$) emitter and a spin-$1/2$ ($\mbox{q}$
or $\overline{\mbox{q}}$) emitter has an 
empty region surrounding it in rapidity of size
$11/6$ and $3/2$, respectively. In order to obtain
this result we note that in terms of the $x$-variables introduced in
Eq.(\ref{e:kinvar}) the factor $\Psi$ is $(x_1^{n_1}+x_3^{n_3})/2$ with
$x_{1,3}=1-k_{\perp}\exp(\pm y)/\sqrt{s}$ and $n_{1,3}$ equal to $2$
or $3$ for $\mbox{q}$($\overline{\mbox{q}}$) and $\mbox{g}$,
respectively. $y_{\mbox{\scriptsize \{max,min\}}}$ are determined from the
energy momentum requirement in Eq.(\ref{e:energycons}).

A note of caution should be issued at this point. For given $s$ and
$k_{\perp}$ there are two definite limits in rapidity
$y_{\mbox{\scriptsize min}} \leq y \leq y_{\mbox{\scriptsize max}}$,
and there is then a depletion of emissions due to helicity conservation,
in regions close to $y_{\mbox{\scriptsize min}}$ and
$y_{\mbox{\scriptsize max}}$. It is in general a poor approximation to
put the factor~$\Psi$ to a unit stepfunction for $y_{\mbox{\scriptsize
min}} +c/2 \leq y \leq y_{\mbox{\scriptsize max}}-c/2$ although it
works when the rapidities and azimuth are integrated out.  A closer
examination provides a $y$-distribution with similarities to a
finite temperature Fermi distribution. We
will nevertheless refer to this feature as ``the excluded region''
around each gluon.

We note that in the process $\mbox{g}\rightarrow
\mbox{q}\overline{\mbox{q}}$, where the spin-$1$ parton emits two spin-$1/2$
partons, that the fermion pair ``prefer'' to be parallel, since there
are no poles in this decay distribution.
However the process $\mbox{g}\rightarrow
\mbox{q}\overline{\mbox{q}}$ is suppressed compared to the 
process $\mbox{g} \rightarrow \mbox{gg}$ and is in general neglected.
The DCM will in this way produce a fan-like set of dipoles, which in
the LLA increases the phase space (the total available effective rapidity range)
for further emissions. However, including the influence from the
polarisation sum (which is essentially the approximation scheme called
Modified LLA) there is in each emission also a depleted region
around an emitted parton, in practice $c=11/6$, because the gluons
completely dominate the process.  At large energies, but not too
large transverse momenta, one may in general neglect the restrictions
but they will be very noticable at the end of the cascades. For example,
with a dipole mass of $3$ GeV the typical rapidity range available
for gluon emission is about $4$ units, and it is then very noticable
to exclude $11/6$ units. 

It is interesting that the average region excluded due to
helicity conservation also occurs in connection with the properties of
the running coupling. To be more precise, we consider a change of
scale in the definition of a field quantum and its interaction. A
change of scale means that the field operator, which has been
normalised to a single quantum at one scale, and the coupling
constant, which likewise has been normalised at the original scale,
will both change. These changes can be read out from the
Callan-Symanzik equations and the $\beta$-function contribution,
stemming from the change in the coupling constant, can be written as
\begin{eqnarray}
\label{e:reabsorb}
- \beta(\alpha_s)\frac{\partial{\cal M}}{\partial \alpha_s}= 
(\frac{11}{6}\frac{N_c \alpha_s}{2\pi} -\frac{2}{3} \frac{n_f \alpha_s}{4 \pi})
\alpha_s\frac{\partial{\cal M}}{\partial \alpha_s}
\end{eqnarray}
where a change in a quantity ${\cal M}$, when the observation scale is
decreased from the level $\kappa= \log(k_{\perp}^2)$ to $\kappa
-d\kappa$, is considered. The decrease accounts for the minus sign on
the left hand side. According to the DCM there is then at this new
scale not only the possibility to emit new gluons but also, at the
next order in the coupling $\alpha_s$, the possibility to reabsorb
already emitted gluons.

The operator $\alpha_s\partial/\partial \alpha_s$ works like a number
operator, i.e. for any function ${\cal M}=\sum \alpha_s^n m_n$ it
provides the number $n$ of possible insertions.  The quantities $N_c
\alpha_s/2\pi$ and $n_f \alpha_s/4 \pi$ are the couplings for
$\mbox{gg}\rightarrow \mbox{g}$ and $\mbox{q}
\overline{\mbox{q}}\rightarrow \mbox{g}$ (and the inverse processes) 
while $11/6$ and $2/3$ corresponds to the effective (generalised)
rapidity ranges available in these reabsorption processes for a given
$\kappa$. It should be noted, however, that this interpretation is
gauge-dependent; in almost all gauges there are contributions to the
$\beta$-function from the vertex corrections. However, for a particular gauge
choice with the propagator given by $-(g_{\mu
\nu}-4k_{\mu}k_{\nu}/k^2)/k^2$, the vertex contributions vanish. 
A closer analysis reveals that the major effect stems from the so-called
Coulomb gluons, i.e. a charged particle like a field quantum in a
non-abelian theory is always accompanied and interacts with its own
Coulomb field. The $11/6$ can therefore be considered as the region
around the gluon containing its accompanying field.
This has been utilized for an approximation of the QCD cascades where
the available phase space for emission is
discretised~\cite{r:discreteQCD}.

\section{A toy model for the end of the cascades}
\label{s:toymodel}

After several gluon emissions there are a set of dipoles with small
masses, and there are in general very many Feynman graphs which may
contribute. The largest diagramatic contribution is chosen according
to coherence conditions in the cascade; in the Dipole
Model~\cite{r:DCM} by an ordering of the gluon emissions in transverse
momentum, and in the Webber-Marchesini model~\cite{r:wm84}, and the
model implemented in JETSET~\cite{r:jetset}, according to a choice of
kinematical variables that fascilitates a strong angular ordering of
the emitted gluons. Results from the cascade models are essentially
equivalent, at least as long as sufficiently hard gluon emission is
considered.

The ordering of emissions in the models will lead to dipoles with
small masses emitting softer gluons. These soft gluons have a
transvere momentum, $k_\perp$, of the same 
order as their emitter and recoils play an
important r\^ole. At present there exists only a minor knowledge of how
the recoils should be distributed among the emitters. 
A sufficiently
large recoil on one of the emitting (soft) gluons will in general
imply that the chosen order is no longer in accordance with the coherence
conditions. 
Emitting soft gluons will evidently lead to a situation
where several, or even very many, paths to the final state are
important, and many different Feynman graphs may contribute and
interfere.

To investigate the emission of soft gluons we propose a toy model with
the following two properties:
\begin{itemize}

\item[I]
We assume that the effective coupling $\bar{\alpha}$ is large enough so
that there is a tendency to emit as many gluons as possible,
essentially with the same $k_{\perp}$.

\item[II]
We assume that the emissions fulfil the requirement of helicity
conservation; this implies that two colour-connected gluons cannot be
closer than a ``distance'' $d=c$.
\end{itemize}

We will use the following combination as the probability for a given
colour-connected multi-gluon state
\begin{equation}
\label{e:manygluem}
P = \prod_1^{n-1} \frac{\alpha \beta}{s_{j,j+1}}
\end{equation}
where $s_{j,j+1}$ is the dipole mass between the colour-connected
gluons $j$ and $j+1$. The factor $\alpha$ corresponds to the product of the 
coupling and the relevant phase space region, and $\beta$ to the
restrictions from helicity conservation, i.e. the requirement of a
suitable distance between the emitted gluons.  Neglecting recoils, we
obtain for any order of the emissions in the DCM, that the product of
factors $1/s_{j,j+1}$ can be written in terms of the invariant dipole
transverse momenta as
\begin{equation}
s_{12}s_{23}\ldots s_{n-1,n}=k_{\perp 2}^2 k_{\perp 3}^2 \ldots k_{\perp n-1}^2 s_{12\ldots n}
\end{equation}
where $k_{\perp j}$ denotes the invariant $k_\perp$ of the dipole from
which gluon $j$ is emitted. Eq.(\ref{e:manygluem}) is therefore a
simple generalisation of Eq.(\ref{e:bremsstrahlung}). 

The dipole mass can be written as
\begin{eqnarray}
\label{e:dipolemass}
s_{j,j+1} & = & k_{\perp}^2 2[\cosh(\Delta y)-\cos(\Delta \phi)] \nonumber \\ 
 & \simeq & k_{\perp}^2 (\Delta y^2 +\Delta \phi^2) [1 + (\Delta y^2 -\Delta
\phi^2)/12].
\end{eqnarray}
For simplicity we have set the transverse momenta of the two gluons to
be identical. $\Delta y$ and $\Delta \phi$ are the differences between
the colour-connected gluons in rapidity and azimuthal angle,
respectively.  
We are now in a position to define precisely what we mean by 
``distance''. We therefore introduce a distance measure, $d$, 
which is related to the dipole mass by 
\begin{equation}
d_{j,j+1}\equiv\sqrt{s_{j,j+1}/k_{\perp}^2} \; \; .
\end{equation}
 When the dipole mass and
the rapidity region are large the azimuthal dependence can be
neglected and $d\simeq \Delta y$.

The emission of soft gluons has thus been reduced to the following problem;
given a certain rapidity range and the full accompanying azimuthal
range $0 \leq \phi <2 \pi$ how are the colour-connected gluons
distributed in phase space in order to obtain a maximum of $P$ in
Eq.(\ref{e:manygluem}), keeping in mind that the gluons cannot be too
close?

From Eq.(\ref{e:manygluem}) we see that the magnitude of $\alpha$
controls the relative probability between different gluon number
states. If $\alpha$ is sufficiently large the number of emitted gluons
will fill the available phase space, and $P$ becomes maximal when the
gluons align along a straight line in phase space. This helix-like
structure is the optimal configuration irrespective of the size of
$\alpha$, or of the number of emitted gluons. For a given multi-gluon
state there are many possible ways to colour-connect the state, where
the helix is only one of the possibilities. It is of course possible
that the sub-optimal configurations are the important ones and swamp the
helix-like contribution, but there are also many contributions close to
a perfect helix.

We have carried out a numerical study to test whether the contributions from
helix-like structures survive the phase space effects. Our program
calculates all possible configurations on a discretized $(y,\phi)$
phase space taking into account that gluons must not be closer than
$c$ to each other. The number of possible configurations grows
factorially with the number of gluons, but the number of gluons is
restricted by the available phase space. We have studied a reasonable
phase space size of three units of rapidity using a closest gluon-to-gluon
distance $c=11/6$ in all the calculations. Since the fluctuations in
dipole $k_\perp$ are limited within a narrow range at the end of the cascades
and the dependence on dipole $k_\perp$ in Eq.(\ref{e:bremsstrahlung}) is
rather weak, we set the transverse momenta of the gluons to be constant.

\begin{figure}[t]
  \hbox{\vbox{ \begin{center}
    \mbox{\psfig{figure=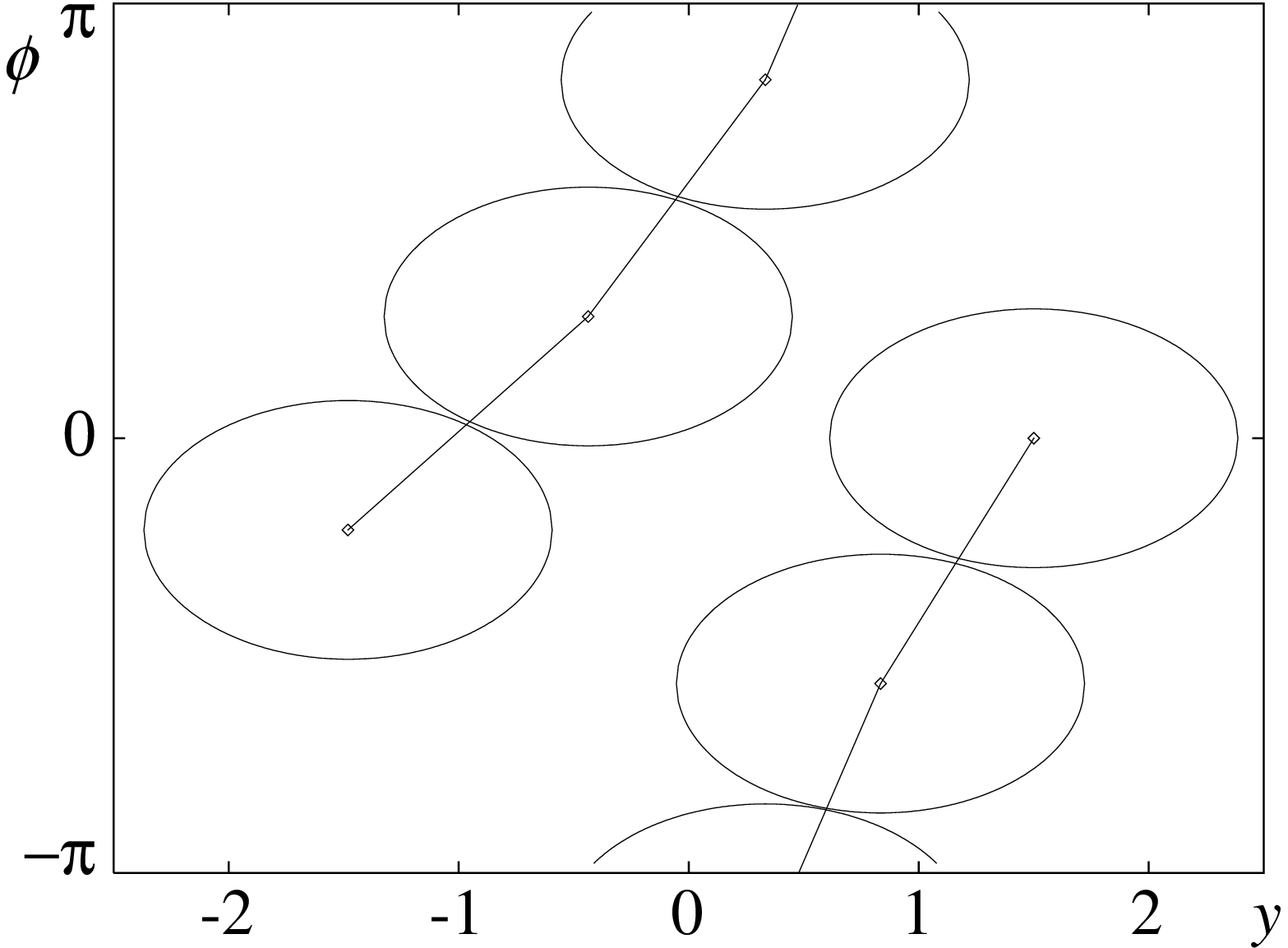,width=8cm}} 
    \hspace{0.2cm}
    \mbox{\psfig{figure=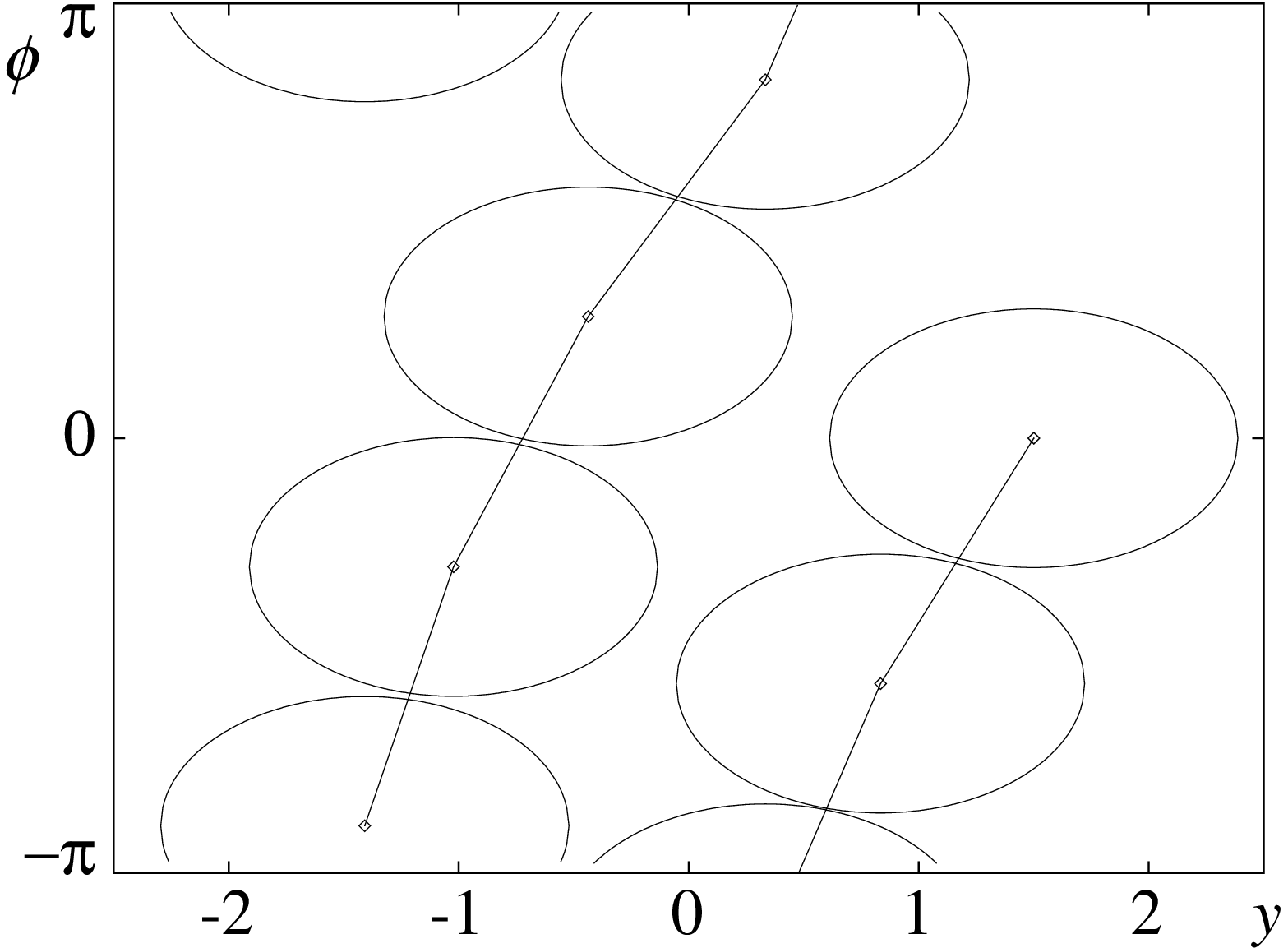,width=8cm}} 
\end{center} }}
\caption{\small The most probable configurations with five and six gluons 
using $c=11/6$. (The cylindrical phase space has been 
mapped onto a plane). The gluon exclusion region for each gluon is 
indicated with the ellipse-like shapes. The line segments show the 
colour field and should form a straight line for a perfect helix. The 
discrepancy is due to the discrete phase space used in our numerical 
analysis.}
\label{f:toy_config}
\end{figure}

In Fig.\ref{f:toy_config} we show the most probable five and six gluon
configurations.
The points corresponding to a given mass correspond to
ellipse-like shapes ($\sqrt{\cosh (\delta y)-\cos(\delta \phi)}$ ) and in
order to minimize the distance between adjacent gluon emissions these
ellipses must be displaced so that they correspond to a helix-shaped
configuration.  The case shown corresponds to the optimal situation
where it is favourable to ``close pack'' the gluons irrespective of the
size of $\alpha$.

\begin{figure}[t]
  \hbox{\vbox{ \begin{center}
    \mbox{\psfig{figure=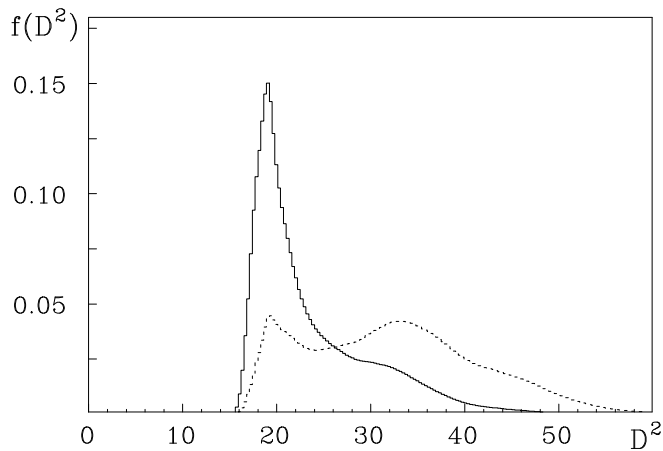,width=10cm}} 
\end{center} }}
\caption{\small The unweighted (dashed line) and 
weighted (solid line) squared length distributions, $f$, of 
configurations with six gluons.}
\label{f:toy_length}
\end{figure}

Taking into account all possible configurations we obtain a
distribution in $D^2 \equiv \sum d_{j j+1}^2$ 
which is very broad, cf. Fig.\ref{f:toy_length}, but
weighting each $D^2$ with the corresponding $P$ from
Eq.(\ref{e:manygluem}) we obtain a large and narrow peak close to the
most probable colour-connected configuration indicating that the gluon
configurations have short strings close to the optimal helix
structure.

Now that we have established that short strings are preferred we
investigate in more detail if they are helix-like in general.  To this
aim we will introduce a new possible observable, ``screwiness''. At
this point it is only a theoretical observable, but later on we will
show how to use it for the final state hadrons. We define
screwiness ${\cal S}$ from the values of $(y_j,\phi_j)$ for the
emitted gluons in accordance with the toy model,
\begin{equation}
\label{e:screwglu}
{\cal S}(\omega) = \sum_{e} P_e 
\left | \sum_j \exp(i(\omega y_j-\phi_j)) \right |^2.
\end{equation}
The first sum is over all the configurations $e$ found in the phase
space and the second goes over the gluons in the configuration. For
$\omega$-values close to zero, screwiness must be small if the gluons
are emitted isotropically in the azimuthal angle. For large values of
$\omega$ the phases should be close to chaotic and then screwiness
only depends on the mean number of emitted gluons.

\begin{figure}[t]
  \hbox{\vbox{ \begin{center}
    \mbox{\psfig{figure=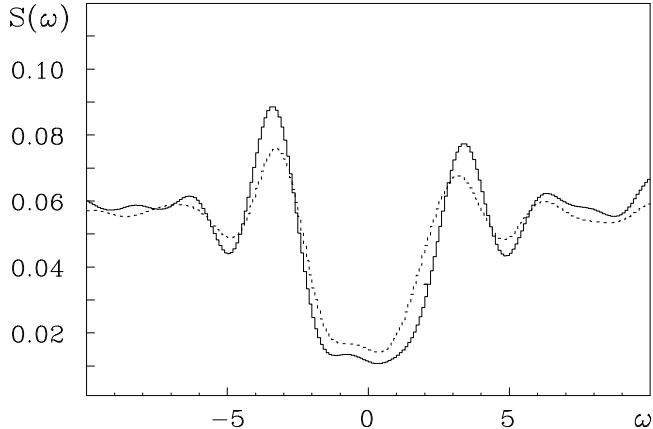,width=10cm}} 
\end{center} }}
\caption{\small Screwiness in the toy model for five (dashed line) and 
six (solid line) gluon states in a rapidity region of three units 
with $c$=11/6.}
\label{f:toy_screwiness}
\end{figure}

In Fig.\ref{f:toy_screwiness} we show the screwiness distribution
including contributions from all configurations with a specific number
of gluons. Two cases are shown, firstly configurations with the maximum
possible number of gluons (in a three unit rapidity phase space this
is six gluons), and secondly those corresponding to five gluon states (the
contributions corresponding to even smaller number of gluons show
similar distributions).  There are two noticable broad peaks with
their mean values close to $\omega = \pm 2\pi/c$. Since the helix
structure has no preferred rotational direction the distributions
should be even. The small apparent asymmetry is due to numerical effects.
We have also analysed the configurations for $c=1.5$ and 3 and these
results are independent of the minimum gluon-to-gluon distance.

From this toy model we see that if we fill the phase space with soft
gluons, which are forbidden to be too close to one another, then
they tend to line up along a helix structure, since the
colour-connection between the gluons prefer to be as
short as possible.

\section{Modelling the helix as an excited string}
\label{s:helixstring}

In order to consider the consequences of the helix-like colour field which
we obtained in Section \ref{s:toymodel} it is necessary to provide
observables in terms of the final state hadrons. A first attempt to
model such a field is to approximate it by the emission
of a set of colour-connected gluons with the same transverse
momentum $k_{\perp}$.
We may then consider the properties of the final state hadrons, as
produced by the Lund string fragmentation model. We very quickly
find that in the competition between increasing the multiplicity versus
increasing the transverse momentum of the hadrons the model
uses the first possibility only. In this section we will be
content with giving the basic argument for why the helix cannot be
described as gluonic excitations on the string field. 

Suppose that a gluon with transverse momentum $k_{\perp}$ is moving
transversely to the constant ($\kappa$) force field, then it is
possible for the gluon to drag out the string field the distance $\ell
= k_{\perp}/2\kappa$ (a gluon experiences twice the force
acting on a quark). On the
other hand, in a quantum mechanical setting such a gluon is only
isolated from the field if the wave-length of the gluon $\lambda
\simeq 2\pi/k_{\perp}$ is smaller than $\ell$ and therefore
\begin{eqnarray}
\label{e:glueformation}
k_{\perp}^2  \geq &  4 \pi \kappa \nonumber \\
\ell_{min}^2 =    &  \frac{\displaystyle \pi}{\displaystyle  \kappa}
\end{eqnarray}
(this is similar to the Landau-Pomeranchuk formation time arguments).
From the first line in Eq.(\ref{e:glueformation}) we obtain the
requirement that a ``real'' gluon must have a transverse momentum
larger than $k_{\perp 0} = 1.6$ GeV. 

We conclude that the helix field cannot be described in terms of a
finite number of gluon excitations on the Lund string. The many
small-$k_{\perp}$ excitations in the model tend to increase the final
state particle multiplicity (with small fluctuations) rather than to
produce transverse momentum for the particles. The interested reader
can find a more thorough investigation of the problems 
associated with the fragmentation
of soft gluons in appendix~\ref{s:softfrag}.

\section{A semi-classical field at the end of the cascades}
\label{s:semi_classical}

We will now consider the possibility that a (semi-)classical colour
field is produced at the end of the perturbative QCD cascades that
cannot be described solely in terms of gluonic excitations on the
Lund Model string field.  The properties of this field should be in
accordance with the toy model that was described in Section
\ref{s:toymodel}. Thus the internal colour quantum number should be
correlated to the external space-time (energy-momentum space)
behaviour so that the colour field has a helix structure, i.e.  the
colour field lines are turning around a spacelike direction, from now
on called the $1$-axis.

\begin{figure}[tb]
  \hbox{\vbox{
    \begin{center}
    \mbox{\psfig{figure=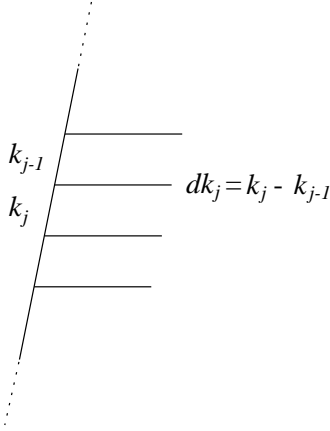,width=4.5cm}}	
    \end{center}
  }}
\caption{\small A current with constant virtuality, $k_j^2=-m^2$, emitting massless field quantas, $dk_j^2=0$.}
\label{f:ladder}
\end{figure}

We may describe the expected field in terms of a wave-packet of
energy-momentum space four-vectors, $k_{\theta}$, corresponding to
the colour current (the index $\theta$ stands for the parameters
describing the wave-packet). We will assume that the vectors
$k_{\theta}$ always have a constant virtuality
$k_{\theta}^2=-m^2$. We further 
assume that the helix colour field is itself emitted
from the current as a continuous stream of gluons $dk$,
colour-connected along each emission vector, $k_{\theta}$. They should
be obtained by differentiating the vector $k_{\theta}$ (we are
generalising the physics picture from a ladder-diagram as in 
Fig.\ref{f:ladder}, where the ``propagator'' vectors $\{k\}_j$ are
emitting the gluons $d k_j=k_j-k_{j-1}$).

The most general description of such a vector is
 (we use
lightcone coordinates along the $01$-direction and transverse coordinates in
the $23$-plane and we do not worry about the initial values):
\begin{eqnarray}
\label{e:kcurr}
k_{\theta} & = & m [cos(\theta)(\exp(y), -\exp(-y),0,0)+  
\sin(\theta)(0,0,\cos(\sigma \phi),\sin(\sigma \phi))] \nonumber\\
& \equiv & m[\cos(\theta)e_1(y) + \sin(\theta)\vec{e}_{\perp 1}(\sigma \phi)].   
\end{eqnarray}
Here $m$ is a constant parameter, $y$ is the rapidity and
$\phi$ the azimuthal angle. We have introduced $\sigma$ as a constant describing
the relative motion in rapidity and azimuth. We 
will put $\sigma=1/2$ later in order to get $\phi$ as 
the azimuthal angle of $dk$. Finally $\theta$ is the
variable describing, on the one hand, the size of the fluctuations in
the longitudinal and transverse parts, and on the other hand,
the properties of the wave-packet.

Assuming that the emitted field quanta $dk$ are massless, we get,
\begin{eqnarray}
dk^2=0 & \Rightarrow & \left(\frac{d\theta}{d\ell}\right)^2 +\left(\frac{1-
cos(2\theta)}{2}\right)= \left(\frac{dy}{d\ell}\right)^2, \nonumber \\ 
\mbox{small}~\theta & \Rightarrow &  \left[\left(\frac{d\theta}{d\ell}\right)^2 +
\theta^2\right]= \left(\frac{dy}{d\ell}\right)^2.
\label{e:massless}
\end{eqnarray}
We have used the differential $d\ell\equiv \sqrt{dy^2+d(\sigma \phi)^2}$. 
Therefore, the assumptions of constant virtuality of $k$ and the
masslessness of $dk$ imply that the variable $\theta$ should fulfil
the pendulum equation according to the first line of
Eq.(\ref{e:massless}). In the limit of small $|\theta|$-values this
becomes a harmonic oscillator equation, assuming that the quantity
$dy/d\ell$ is a (small) constant along each vector $k_{\theta}$. For
consistency we will then make the change $d\ell \rightarrow \sigma
d\phi$.  This is the second line of Eq.(\ref{e:massless}) and using
the notation $dy/d\phi\equiv \tau$ we obtain as a classical
description (again neglecting the boundary values):
\begin{eqnarray}
\label{e:harmosc}
\theta= \frac{\tau}{\sigma} cos(\sigma \phi)
\end{eqnarray}
If we choose $\sigma=1/2$ to make $\phi$ the azimuthal angle
of $dk$, then we find that the field emission
vectors $dk/d\phi$ and the corresponding current vector $k_{\theta}$ are (in
the approximation of small oscillations):
\begin{eqnarray}
\label{e:kresult}
\frac{dk}{d\phi}= m \tau [e_0(y) + \vec{e}_{\perp 0}(\phi)] \nonumber \\
k_{\theta}= m [e_1(y) +\tau (\vec{e}_{\perp 1}(\phi)+ \vec{e}_{\perp 1}(0))]
\end{eqnarray}

Here we have introduced the vectors $e_0 = de_1/dy$ and
$\vec{e}_{\perp 0}= d\vec{e}_{\perp 1}/d\phi$ (note that all the
occurring vectors are orthogonal). 
We may evidently use the quantity
$\tau$ (together with suitable boundary values) to label the wave
packet for the current. That is to say, we may assume that there is a
distribution $h(\tau)$ which
describes the occurrence of the different current lines, each with a
well-defined direction $\tau$. This distribution, $h(\tau)$, 
should be similar to a Gaussian.
A single current line with fixed $\tau$ may also be described in the
transverse plane. The current turns around the $1$-axis with the
azimuthal angle and the corresponding field quanta are emitted
transversely to the current at every emission point according to
Eq.(\ref{e:kresult}). There is one reasonable restriction: the field
energy emitted by the current in a small angular segment should not
exceed the energy which should be available in the Lund Model string.
If we use the string radius as calculated in
Eq.(\ref{e:glueformation}), $\ell_{min}= \sqrt{\pi/\kappa}$, 
then we find that
\begin{eqnarray}
m\tau \leq \kappa\ell_{min} \simeq 0.8~\mbox{GeV} 
\label{e:restrict}
\end{eqnarray}  
It is interesting to note that these fields have similarities to those
studied in connection with dimensional reduction in~\cite{r:nielsen}.

\section{Fragmentation and screwiness}
\label{s:fragmentation}

We have in the previous section described the emission of a continuous
stream of colour-connected gluons having the property that the
azimuthal angle of the stream is proportional to the rapidity, i.e. it
is of a helical character. As previously discussed we cannot implement
this as individual gluonic excitations of the Lund string. We will in
this section instead describe a possible way to take the transverse
properties of the continuous helix into account whilst keeping the
major properties of the Lund fragmentation model. In order to do
this we will begin by presenting a few relevant parts of the
Lund model. This model has been described several times and a recent
investigation can be found in~\cite{r:Peteretal}.

\subsection{The Lund fragmentation process}
\label{s:Lundfrag}

\begin{figure}[tb]
  \hbox{\vbox{
    \begin{center}
    \mbox{\psfig{figure=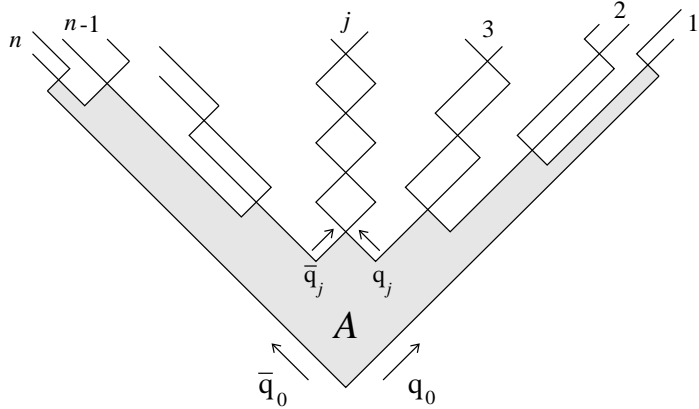,width=9.2cm}}	
    \end{center}
  }}
\caption{\small The break-up of a Lund string.}
\label{f:lunddecay}
\end{figure}

The following (non-normalised) probability to produce a set of 
hadrons has been derived using semi-classical
arguments in \cite{r:BoS}
\begin{eqnarray}
\label{e:Lundprob}
dP(\{p\}_j;P_{tot})= \prod_1^n N_j dp_j \delta(p_j^2-m_j^2) \delta (\sum p_j -
P_{tot}) \exp(-b A).
\end{eqnarray}
Here $N_j$ are normalisation constants, $A$ 
the decay area, cf. Fig.\ref{f:lunddecay}, and $b$ a 
basic colour-dynamical parameter; from comparison to
experimental data we know that $b \simeq 0.6$ GeV$^{-2}$ if the area $A$ is
expressed in energy-momentum space quantities. 

The constant force field spanned between a
colour-$3$ quark and a colour-$\overline{3}$ anti-quark is a simple mode
of the massless relativistic string. The process has been generalised 
into a situation with multigluon emission in \cite{r:TS1}
using the Lund interpretation that the gluons are internal excitations on the
string field.

The area decay law in Eq.(\ref{e:Lundprob}) can be implemented as an
iterative process, in which the particles are produced in a stepwise
way ordered along the positive (or negative) light-cone. If a set of
hadrons is generated, each one takes a fraction $z$ of the remaining
light-cone component $E+p_l$ (or $E-p_l$, if they are generated along
the negative light-cone), with $z$ given by the distribution
\begin{equation}
f(z)=N\frac{(1-z)^a}{z} \exp(-bm_\perp^2/z) .
\label{e:fragfun}
\end{equation}

The parameters $N$, $a$ and $b$ are related by normalisation, leaving
two free parameters. The transverse mass parameter in the fragmentation function
is $m_{\perp}^2= m^2 +
\vec{p}_{\perp}^{\,2}$, with the transverse momentum obtained as the sum of
the transverse momenta stemming from the $\mbox{q}$ and
$\overline{\mbox{q}}$ particles generated at the neighbouring
vertices, $\vec{p}_\perp= \vec{k}_{\perp 2}-\vec{k}_{\perp 1}$. In the
Lund model a $\mbox{q}\overline{\mbox{q}}$-pair with transverse
momenta $\pm k_\perp$ is produced through a quantum mechanical
tunneling process. It results in a Gaussian distribution
for the transverse momenta
\begin{equation}
\label{e:pttunnel}
d^2k_{\perp} \exp(-\pi k_{\perp}^2/\kappa).
\end{equation}
The whole process is implemented in the Monte Carlo program JETSET~\cite{r:jetset}.

\begin{figure}[tb]
  \hbox{\vbox{ \begin{center}
	\mbox{\psfig{figure=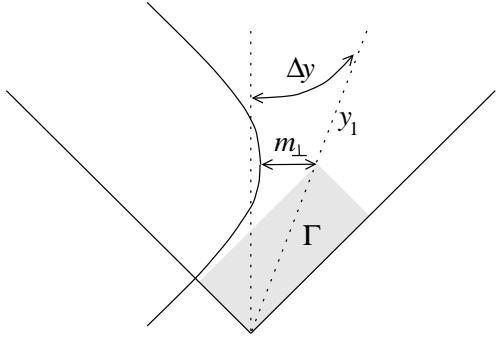,width=6.6cm}} \end{center} }}
\caption{\small The longitudinal energy scale in the Lund model is $\langle \Gamma \rangle$. The figure shows the production of a particle with transverse mass $m_\perp$.
The difference in rapidity between the constituent vertices has to be related to the $\Gamma$ of one of the vertices in order for the vertices to be specified.}
\label{f:deltay}
\end{figure}

Consider the production of a particle with transverse mass
$m_\perp$. Given that one vertex has the rapidity $y_1$, the rapidity
difference $\Delta y$ is not enough to specify the position of the
other vertex. One must also know the proper-time of the first
vertex. This is shown in energy-momentum space in Fig.\ref{f:deltay}
where the first vertex is specified by $\Gamma$ which is the squared
product of the proper-time and $\kappa$.
Of course there are two solutions in this case, but one is
strongly favoured by the area dependence in Eq.(\ref{e:Lundprob}).  In
the Lund model the vertices, on average, lie on a hyperbola given by a
typical $\Gamma$.  That is to say, the steps in rapidity in the particle
production are related to the scale $\langle \Gamma \rangle$ as given
by the model.
There is a similar situation in the transverse momentum generation.
The squared transverse momentum of a particle is not only
given by the azimuthal angle $\Delta \phi$ between the break-up points
that generate the particle.  The lengths of the transverse momenta of
the $\mbox{q}$ and the $\overline{\mbox{q}}$ that make up the particle
are also needed. In the tunneling process in Eq.(\ref{e:pttunnel})
these sizes are given by the scale $\kappa/\pi$.
  
Thus the Lund fragmentation model provides two different energy
scales; one longitudinal to relate to the rapidity difference
between vertices and one transverse to relate to their
difference in azimuthal angle.

\subsection{A modified fragmentation process with screwiness}

The main idea in the screwiness model is that the transverse momentum
of the emitted particles stems from the piece of screwy gluon field
that is in between the two break-up points producing the particle.
Therefore we begin by summing up the transverse momentum that is
emitted between two points along the field line,
cf. Eq.(\ref{e:kresult}):
\eqbe
\int^2_1 \frac{\vec{dk}_\perp}{d\phi}d\phi = \vec{k}_{\perp 2}- \vec{k}_{\perp 1} = m\tau [\vec{e}_{\perp 1}(\phi_2)- \vec{e}_{\perp 1}(\phi_1)]. 
\label{e:kt_screw}
\eqen
\begin{figure}[tb]
  \hbox{\vbox{
    \begin{center}
	\mbox{\psfig{figure=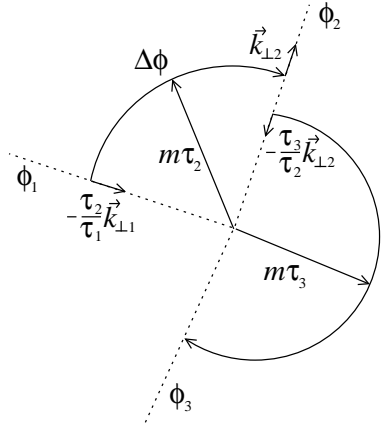,width=5cm}}
    \end{center}
  }}
\caption{\small A $\mbox{q}\overline{\mbox{q}}$-pair 
is produced in a break-up point 
with azimuthal angle $\phi$. 
The figure illustrates 
how the screwy gluon field between $\phi_1$ and $\phi_2$ is 
associated with the transverse 
momentum ($-k_{\perp 1},k_{\perp 2}$) 
of the quarks produced at the two break-up points. This 
association has the property 
that the produced transverse 
momentum is conserved locally in each break-up point.} 
\label{f:deltaphi}
\end{figure}
We note that the quantity $m\tau$ also occurs here. We will always
consider the parameter $m$ to be a suitable fixed mass parameter but
according to the assumed wave function for the current the direction
$\tau$ may vary between the different break-up points.  To keep the
presentation clear we will start off keeping $\tau$ fixed.  In the
end we will present the generalisation to the case of a varying $\tau$.

If we associate $\pm \vec{k}_{\perp i}$ with the transverse momenta 
of the $\mbox{q}\overline{\mbox{q}}$-pair 
produced at vertex $i$, the transverse momenta of 
the produced particles are given by Eq.(\ref{e:kt_screw}).
The corresponding squared transverse momentum is then
\begin{equation}
p^2_{\perp i}= 2m^2\tau^2 [1-\cos(\Delta \phi)]
\end{equation}
where $\Delta \phi = \phi_i-\phi_{i-1}$.  Since $\Delta \phi$ is
proportional to the rapidity difference between vertices $\Delta y$,
it can be written as a function of the particle's light-cone fraction
$z$
\eqbe
\Delta \phi = \frac{\Delta y}{\tau}= 
\frac{1}{2\tau}\log\left(\frac{z+m^2_\perp/\Gamma} {z(1-z)} \right)
\label{e:deltaphi}
\eqen
where $\Gamma$ is defined as in Fig.\ref{f:deltay} and with respect to
the previous break-up point $i-1$. Taken together this means that we
can write the transverse momentum of a particle as a function of $z$
and $\tau$
\eqbe
p^2_{\perp}(z) = 2m^2\tau^2 
\left[1-\cos\left(\frac{\Delta y(z)}{\tau_{i}}\right) \right].
\label{e:ptscrew}
\eqen
As explicitly manifested in Eq.(\ref{e:ptscrew}) this means that the
transverse and longitudinal components are connected in this
model. Inserting $p^2_\perp(z)$ in the Lund fragmentation function gives
\eqbe
f(z)=N \frac{(1-z)^a}{z}
\exp\left(-\frac{b}{z}\left(m^2_h+p^2_\perp(z)\right)\right).
\label{e:screwfun}
\eqen
In this way Eq.(\ref{e:screwfun}) gives 
the distribution of light-cone fractions
for a given direction $\tau$.

This model keeps the longitudinal properties of the
ordinary Lund fragmentation model, but the azimuthal properties are
changed. Rapidity differences are still related to $\langle \Gamma
\rangle$ but steps in the azimuthal angle are now correlated with steps
in rapidity. The azimuthal angles are no longer related to 
$\kappa/\pi$ as given by the tunneling process, but instead to
$m^2\tau^2$ as given by the screwy gluon field.

When going from one vertex to the next in the case of varying $\tau$
one has to keep in mind that the transverse momentum produced at the
first vertex has been specified by the previous step. In order to
conserve the transverse momenta generated at each vertex we therefore
modify the association in Eq.(\ref{e:kt_screw}),
as follows
\eqbe
\int^i_{i-1} \frac{\vec{dk}_\perp}{d\phi}d\phi= \vec{k}_{\perp i}-\frac{\tau_{i}}{\tau_{i-1}}\vec{k}_{\perp i-1}
\eqen
where $\tau_i$ denotes the direction between break-up points $i-1$ and
$i$, cf. Fig.\ref{f:deltaphi}.  The transverse momenta of the produced 
particles are then given by
\eabe
\vec{p}_{\perp i} &=& \vec{k}_{\perp i}-\vec{k}_{\perp i-1}= m\left[\tau_i\vec{e}_{\perp 1}(\phi_i)-\tau_{i-1} \vec{e}_{\perp 1}(\phi_{i-1})\right],\nonumber \\
p^2_{\perp i} &=& m^2 \left[\tau^2_{i} + \tau^2_{i-1} -2\tau_{i}\tau_{i-1}\cos(\Delta \phi) \right].
\label{e:finalpt2}
\eaen
The $p_{\perp}^2$ given by Eq.(\ref{e:finalpt2}) can then be put into
the fragmentation function. Varying $\tau$ results in larger
variations in the emitted transverse momenta of the particles.  We
have used a Gaussian distribution of $\tau$-directions and we have
approximated $m_\perp^2$ in Eq.(\ref{e:deltaphi}) with $m_\perp^2
\simeq m_h^2+\langle p_{\perp min}^2\rangle =
m_h^2+2m^2\sigma^2_\tau$. Where $m_h$ is the hadron mass and
$\sigma_\tau$ denotes the width in the distribution of
$\tau$-directions. The equations can be solved iteratively without
this approximation, but we find that our results are unaffected by this
approximation.

\section{Is screwiness observable?}
\label{s:results}

In this section we will address the question 
of whether introducing a
correlation between $y$ and $\phi$ of the string break-up vertices has
observable consequences for the produced particles. There are two
processes which in principle can destroy such a correlation. Firstly,
there is the initial particle production 
and secondly, there are resonance decays.
The initial particle production spoils things
because even if the vertices 
lie on a perfect helix the produced particle will
usually not lie on the line between its two
constituent vertices in the $(y,\phi)$-plane.
The particle production fluctuations are mainly in rapidity,
i.e. a particle is produced with an azimuthal angle which roughly
corresponds to the average angle of its constituent vertices, while
its rapidity is distributed with width unity around the average of
the vertices.

To study the consequences of the screwiness model we have generated
events with three different values $\langle
\tau \rangle = 0.3 ,0.5 ~\mbox{and}~ 0.7$. For each value we have
tuned the parameters of the model to agree with the multiplicity,
rapidity and transverse momentum distributions of default
JETSET. In this way we can study the correlations introduced by the
model as compared to the ordinary Lund string model. 
We have tuned $m$ to get the default average $p_\perp$ of the produced
particles, utilizing the fact
that the product $m\tau$ is the important
factor. The parameter $b$ has been changed from the default JETSET
value to tune the multiplicity, and $\sigma_\tau$ has been tuned to get
the final charged $p_\perp$ fluctuations. Tuning with different
$\langle \tau \rangle$ values results in the parameter values shown in
Table~\ref{t:parameters}.
\begin{table}[ht]
\begin{center}
\begin{tabular}{|c|c|c|c|} \hline
  $\langle \tau \rangle$ & 0.3 & 0.5 & 0.7 \\ \hline
$m$ & 1.0 & 0.71 & 0.61 \\
$b$ & 0.64 & 0.68 & 0.7 \\
$\sigma_\tau$ & 0.2 & 0.3 & 0.35 \\ \hline
\end{tabular}
\caption{\small Parameter values. The model has been tuned to the multiplicity and charged final $p_\perp$ distributions of default JETSET ($b=0.58$).}
\label{t:parameters}
\end{center}
\end{table}
We note in particular that to get the multiplicity distributions of default
JETSET only minor changes of the $b$-parameter are needed.  We also
note that the restriction in Eq.(\ref{e:restrict}) is satisfied for
all the cases since in this model 
only a fraction of the energy available in the
Lund string is used to produce transverse momenta.

We have generated pure $\mbox{q}\overline{\mbox{q}}$ events
and the particles in the central rapidity plateau have been
included in the analysis. The plots shown are for four units of rapidity,
but the qualitative results for observable screwiness
are unaffected for values as low as approximately three units of central
rapidity.  
We have analysed the properties of the generated events by means of
the screwiness measure, defined in Eq.(\ref{e:screwglu}). Here, the
second sum in the measure instead goes over the hadrons or over the
break-up vertices.  The weight $P_e$ is of course unity for all
events.

\begin{figure}[t]
  \hbox{\vbox{ \begin{center}
    \mbox{\psfig{figure=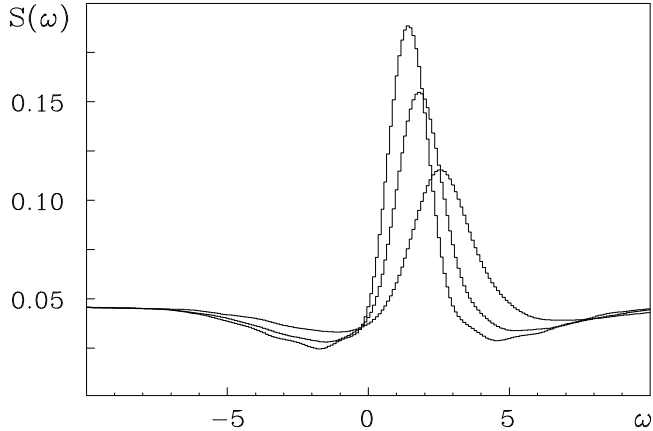,width=10cm}} \end{center} }}
\caption{\small Screwiness for the string break-up vertices. The 
three curves shown are 
for $\langle \tau \rangle = 0.3,~0.5~\mbox{and}~0.7$, respectively. There 
is a clear peak at $\omega \simeq 1/\langle \tau \rangle$ in all the cases.}
\label{f:screw_vertex}
\end{figure}

\begin{figure}[t]
  \hbox{\vbox{
    \begin{center}
    \mbox{\psfig{figure=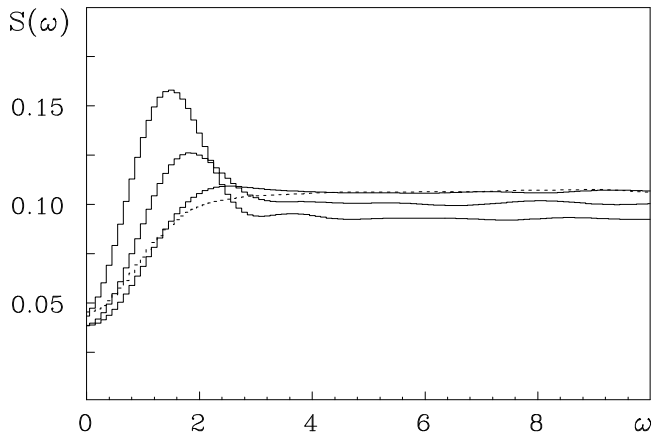,width=10cm}}
    \end{center}
  }}
\caption{\small Screwiness for the directly produced particles. The three solid curves are for  $\langle \tau \rangle = 0.3,~0.5~\mbox{and}~0.7$, respectively. The peak decreases as $\langle \tau \rangle$ is reduced. For $\langle \tau \rangle=0.3$ the peak has vanished due to the fluctuations in particle production. The screwiness for default JETSET (dashed line) has been included for comparison.}
\label{f:screw_particles}
\end{figure}

\begin{figure}[t]
  \hbox{\vbox{ \begin{center}
    \mbox{\psfig{figure=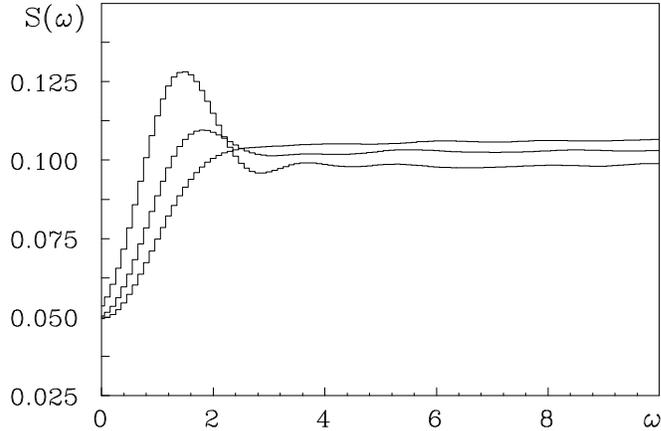,width=10cm}} 
  \end{center} }}
\caption{\small Screwiness for the final particles ($\pi^0$'s are set stable). The three curves shown are for $\langle \tau \rangle = 0.3,~0.5~\mbox{and}~0.7$, respectively. For not too small $\langle \tau \rangle$-values there is a peak at $\omega \simeq 1/\langle \tau \rangle$.}
\label{f:screw_final}
\end{figure}

In Fig.\ref{f:screw_vertex} the screwiness for the break-up vertices
is shown. It has a clear peak for the different values of $\langle
\tau \rangle$, and the $\omega$-values for the peaks correspond to the
average $\tau$ values used.
The screwiness for the initially produced particles is shown in
Fig.\ref{f:screw_particles}. We note that the peak vanishes for small
values of $\tau$. A helix where the windings are separated by two units of
rapidity corresponds to $\tau = 1/\pi$. The vanishing of the signal for
small $\tau$ values is therefore in agreement with our findings for the
rapidity fluctuations in the particle production.
For comparison we have included the screwiness for the initial particles
produced by default JETSET in Fig.\ref{f:screw_particles}. As expected no
signal is found in this case.  The screwiness is
further diluted by resonance decays, but it is still visible for not
too small $\tau$ values as shown in Fig.\ref{f:screw_final}.

To try to enhance the signal we have investigated how the screwiness
measure depends on multiplicity and the transverse momentum of the
particles. Selecting events with large initial multiplicity enhances
the signal. However, analysing events with different final
multiplicities separately does not give an enhancement of the
signal. The influence of resonance decays on the multiplicity is
too large. 

\begin{figure}[t]
  \hbox{\vbox{
    \begin{center}
    \mbox{\psfig{figure=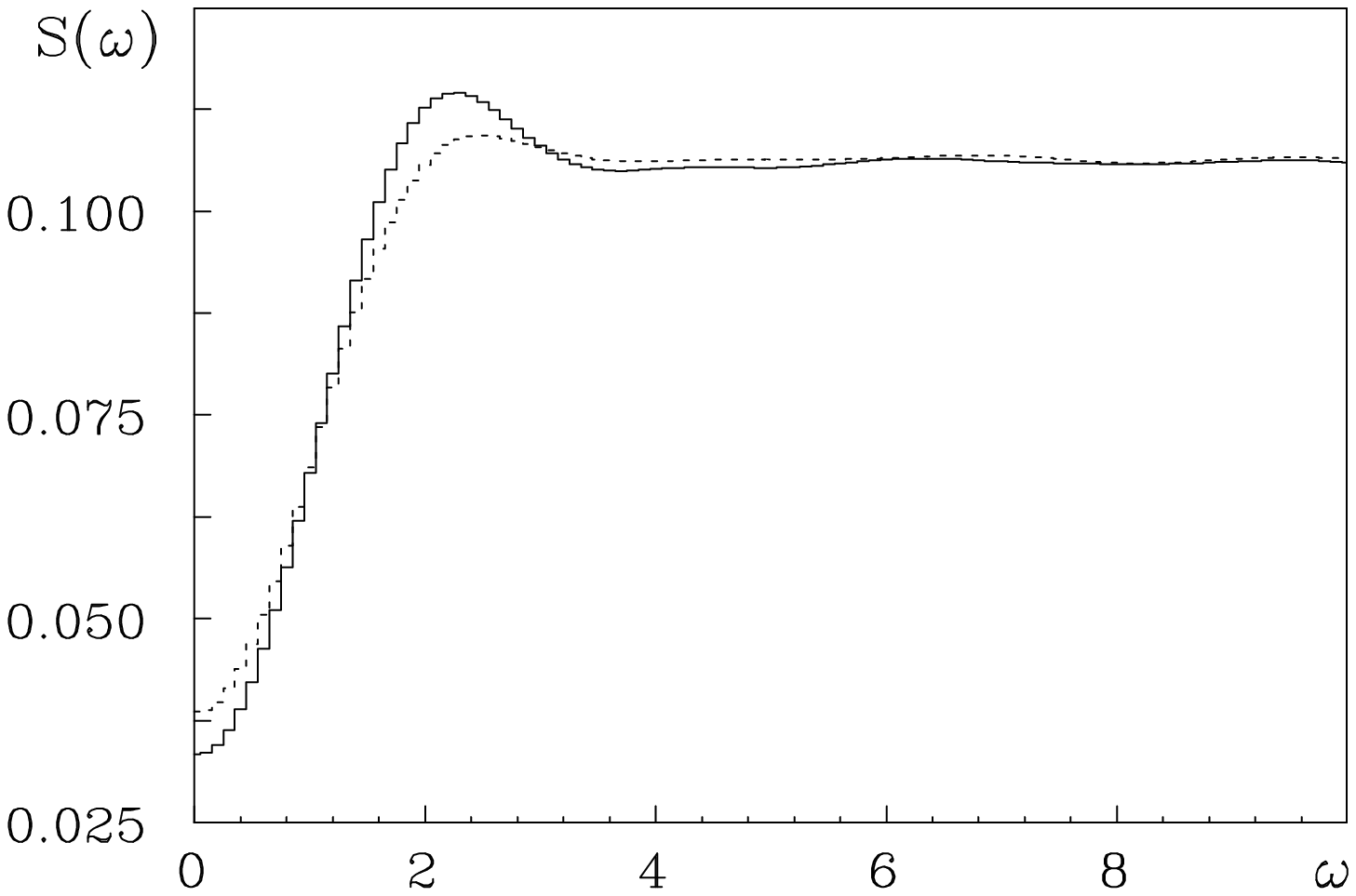,width=8cm}}
    \mbox{\psfig{figure=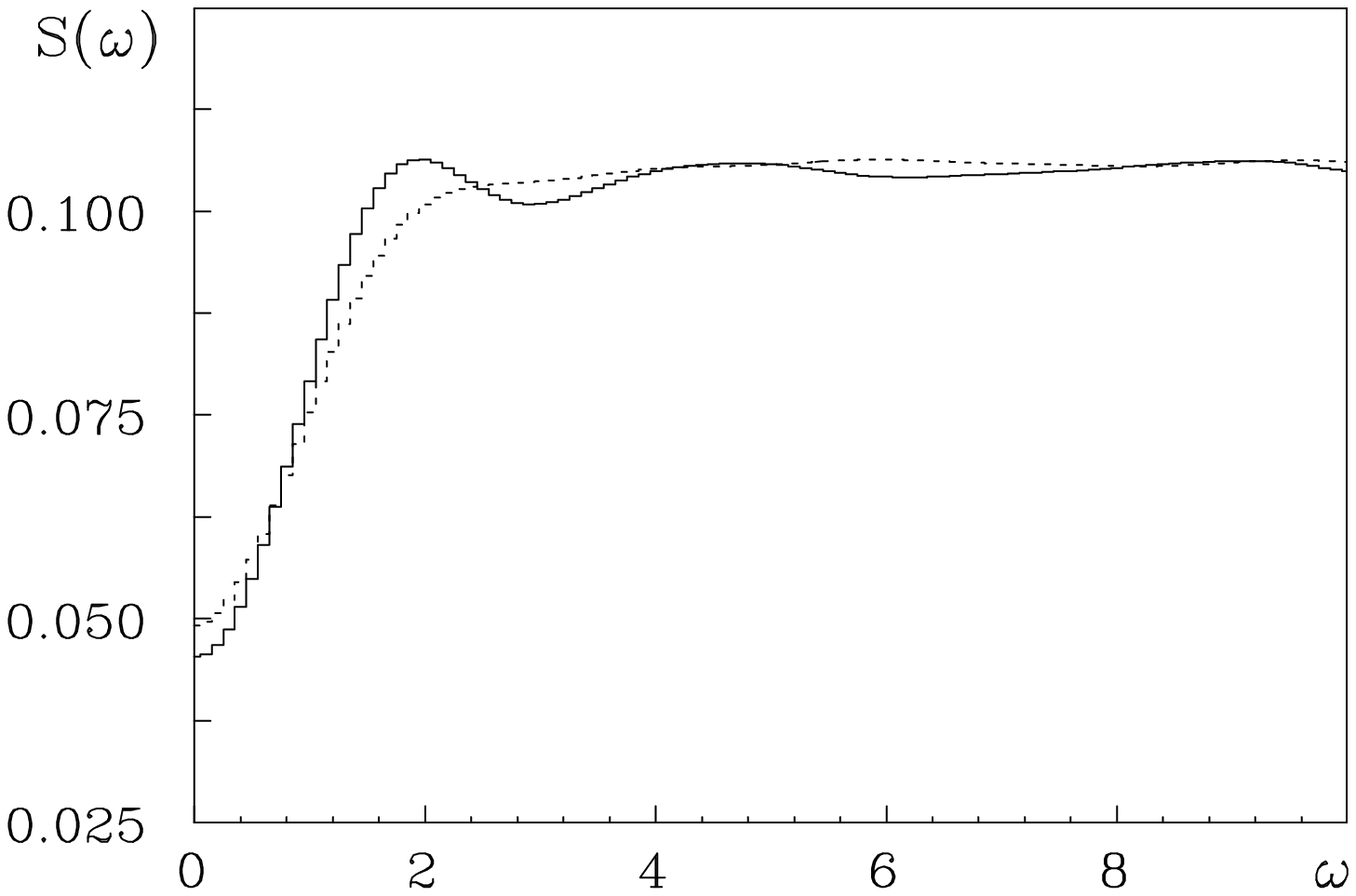,width=8cm}}
    \end{center}
  }}
\caption{\small Screwiness for $\langle \tau \rangle=0.3$. To enhance the signal events where $\langle p_\perp^2 \rangle$ is large have been selected (solid lines). We have included the corresponding curve with no event selection (dashed lines) to indicate the improvement. Left) Initially produced particles. $\langle p_\perp^2 \rangle>0.3~\mbox{GeV}^2$. Right) The final particles. $\pi^0$'s are set stable. $\langle p_\perp^2 \rangle>0.25~\mbox{GeV}^2$.}
\label{f:screw_max}
\end{figure}

Selecting events where $\langle p^2_{\perp} \rangle$ is large enhances
the signal when decays are not included. This is shown for $\langle
\tau \rangle =0.3$ in the left part of Fig.\ref{f:screw_max} where
events with $\langle p_\perp^2
\rangle>0.3$~GeV$^2$ for the initial particles have been selected. As shown in
the figure this event selection results in the signal surviving
particle production even for small $\langle \tau \rangle$-values.
This event selection is also profitable when it comes to decreasing
the effects of resonance decays since events with many decay products
are not likely to be selected. In the right part of
Fig.\ref{f:screw_max} we show the 
screwiness for the final state particles in
events where $\langle p_\perp^2 \rangle>0.25$~GeV$^2$. The curves
shown are for $\langle \tau \rangle=0.3$ to show that with event
selection a signal can be obtained {\it even for this case}. Using the same
event selection of course enhances the signal for larger $\langle \tau
\rangle$-values, but in those cases it was clearly visible in the total sample.

A total of 50000 $\mbox{q}\overline{\mbox{q}}$ events have been used
in the analysis, except in the event selection analysis in
Fig.\ref{f:screw_max} where 250000 events are analysed. To be
able to observe screwiness for such a small $\langle \tau
\rangle$-value one needs to increase the number of 
events by a factor of about five compared to the
larger values. Since we have only used positive $\langle \tau
\rangle$-values, events with a preferred rotational direction are
generated.  We could have included both rotational directions in the event
generation which would add a signal for negative $\omega$, but
reduce the statistics by a factor of two.

The effects on the screwiness from hard gluons stemming from the
parton cascade will be investigated in future work. However, since
only a fairly small number of events are needed for the results in this
paper we expect that investigations of experimental data,
in which hard gluon activity is excluded, can be profitable.

\begin{figure}[t]
  \hbox{\vbox{
    \begin{center}
    \mbox{\psfig{figure=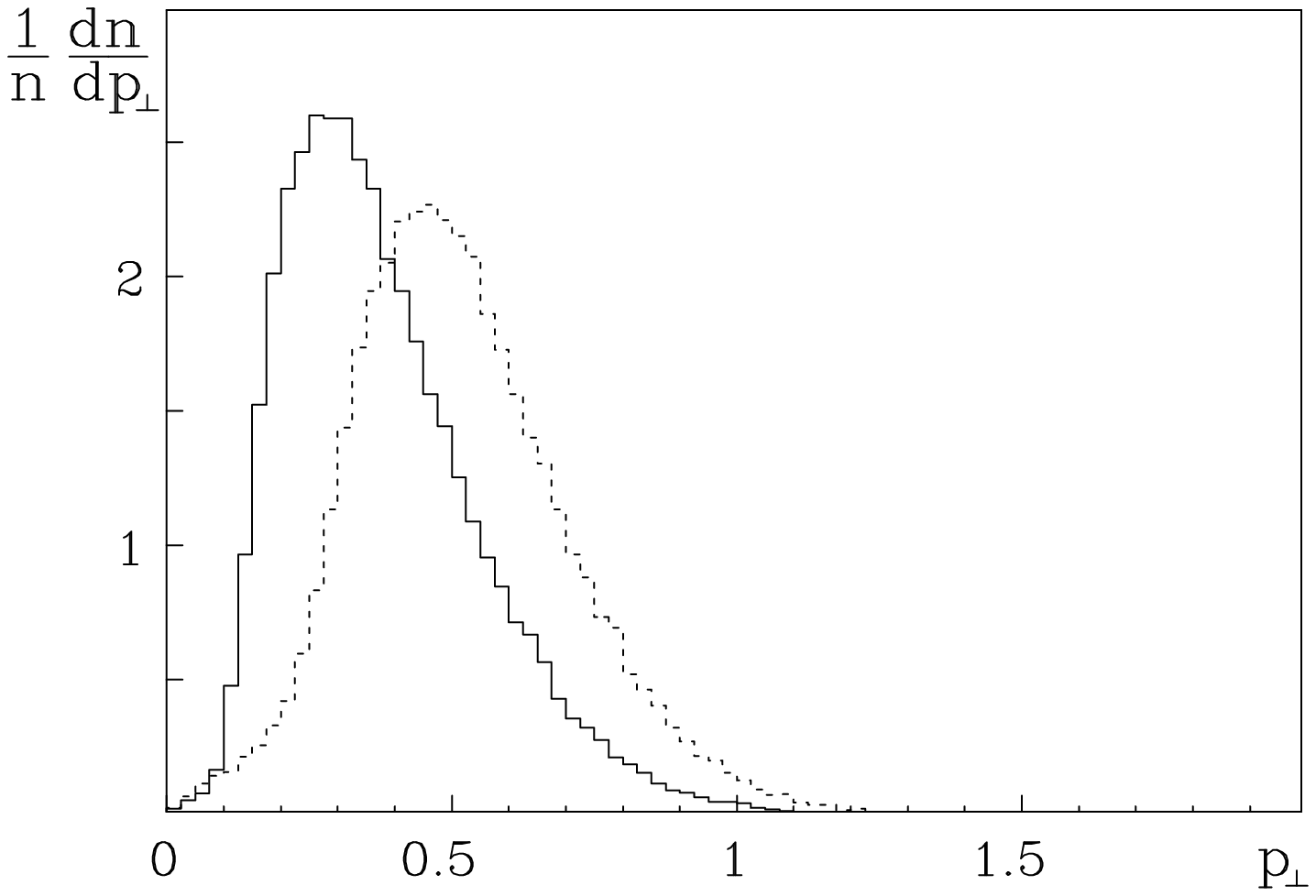,width=8.0cm}} 
    \mbox{\psfig{figure=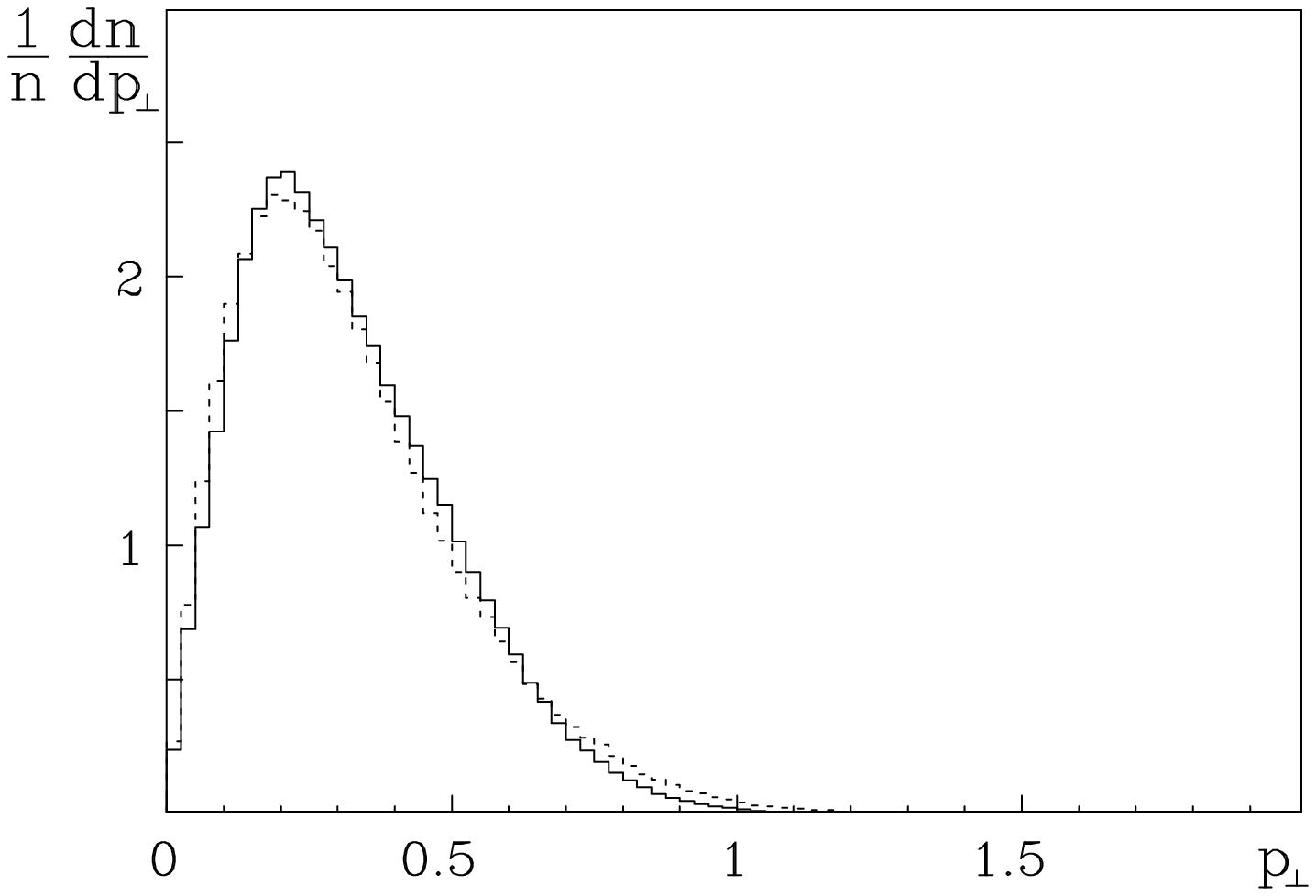,width=8.0cm}}
    \end{center}
  }}
\caption{\small Left) The $p_\perp$ (GeV) distributions for the 
directly produced $\pi$'s (solid) and $\rho$'s (dashed). 
The curves shown are for $\tau=0.5$. 
Right) The $p_\perp$ (GeV) 
distributions for the all final pions 
(solid), $\pi^0$'s are set stable, as compared with default JETSET (dashed).}
\label{f:screw_pt}
\end{figure}

A specific property of our model is that $\langle p_{\perp}^2 \rangle$
for directly produced pions is smaller than $\langle p_{\perp}^2
\rangle$ for heavier particles. This feature
appears to be in agreement with experimental data on two-particle
correlations~\cite{r:tasso}. A model for correlations in $p_\perp$ in
the string hadronization process with similar consequences was
introduced in~\cite{r:jimpt}. The $p_\perp$ for directly produced
pions and $\rho$'s are shown for the screwiness model in
Fig.\ref{f:screw_pt} and the distributions are clearly different. In
the figure we also show the $p_\perp$ distribution of the final
pions and compare it to the default JETSET distribution.
As seen the secondary pions wash out the differences. The $\langle
p_\perp \rangle$ for various flavours at the initial production level
depend on the screwiness parameters, but the qualitative difference
remains.

\section {Conclusions}
It is perhaps surprising that such an ordered structure as a helix
could emerge at the end of the QCD cascade. However, when we consider
the constraint imposed by helicity conservation, we see that purely
random configurations of gluons are disfavoured. This is because the
exclusion region around each gluon restricts the maximum number
of allowed gluons.  Instead we see that the gluons can achieve the maximum
concentration by close packing themselves into the form of a helix.
The fragmentation of this screwy field has consequences for the final
state particles. Although the fragmentation cannot be described in
terms of gluon excitations of the Lund string, we have instead
modified the Lund fragmentation scheme. If the winding is within
reasonable limits then we expect ``screwiness''
to be an observable feature of the QCD cascade.

\vspace{1cm} {\noindent \large \bf Acknowledgments}\vspace{0.5cm} \\
We thank Patrik Ed\'en for very many valuable discussions. This work
was supported in part by the EU Fourth Framework Programme `Training
and Mobility of Researchers', Network `Quantum Chromodynamics and the
Deep Structure of Elementary Particles', contract FMRX-CT98-0194 (DG
12 - MIHT).

\appendix
\section{Problems with fragmenting soft gluons}
\label{s:softfrag} 

In section~\ref{s:helixstring} we claimed that the helix colour-field
cannot be implemented as an excited string, since gluons softer than
$k_{\perp 0}=1.6$~GeV cannot be considered as excitations of the
string.  

To illustrate the problems with fragmentation of soft gluons we have
investigated JETSET fragmentation of parton configurations with soft
gluons emitted according to the Dipole Cascade Model as implemented in
the ARIADNE Monte Carlo~\cite{r:ariadne}. The allowed $k_\perp$ range
for emissions from the colour dipoles is normally between an upper
value, given by phase-space limits, and a lower infra-red cut-off,
$k_{\perp c}$. We have instead used a small maximum allowed $k_\perp$
value (denoted $k_{\perp \mbox{\scriptsize max}}$) to restrict
the hardness of the emitted gluons. This soft cascade has been
applied to $\mbox{q}\overline{\mbox{q}}$-dipoles oriented along the
$z$-axis.  The soft gluons have a negligible impact on the event
topology and for our purposes it therefore makes sense to define
rapidity with respect to the $z$-axis. We have analysed
the resulting 
hadrons in the central rapidity plateau of the events.  To emphasize
the features of fragmentation of soft gluons we have not included
resonance decays in our analysis.

\begin{figure}[tb]
  \hbox{\vbox{
    \begin{center}
    \mbox{\psfig{figure=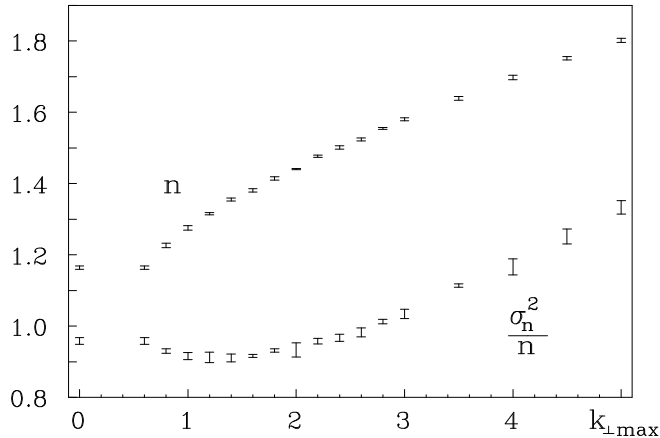,width=10cm}}
    \end{center}
  }}
\caption{\small The multiplicity in 
central rapidity per unit of rapidty $n$ and 
the corresponding variance $\sigma_n^2$ 
depends on the upper cut-off in the 
cascade $k_{\perp \mbox{\scriptsize max}}$~(GeV) 
as shown. Default JETSET has been used for 
fragmentation and $\sigma_n^2/n$ does not start 
to increase until $k_{\perp \mbox{\scriptsize max}}$ is roughly 1.6~GeV}
\label{f:softgluefrag}
\end{figure}

In Fig.\ref{f:softgluefrag} we show how the average and the squared width 
of the central multiplicity distribution depend on
$k_{\perp \mbox{\scriptsize max}}$. 
The effect of the soft gluons is an increase of the average multiplicity
while the multiplicity fluctuations remain constant or even decrease
until $k_{\perp \mbox{\scriptsize max}}$ 
is above $k_{\perp 0}$. The $\langle p_\perp
\rangle$ with respect to the $z$-axis of the hadrons only increases
from 0.46 GeV for a flat string with no gluon excitations to 0.56 
GeV for
$k_{\perp \mbox{\scriptsize max}}=3$ GeV. Changing the generated $\langle
p_\perp \rangle$ by such a small factor has a minor effect ($\sim
5\%$) on the average multiplicity in pure
$\mbox{q}\overline{\mbox{q}}$ events whilst adding the soft gluons
increases the average multiplicity by roughly $40\%$, as shown in the
figure. As mentioned in section~\ref{s:helixstring}, we find that the
soft gluons essentially only increase the hadron multiplicity. The
number of gluons per rapidity unit varies from 0.25 for $k_{\perp
\mbox{\scriptsize max}}=1$ GeV to 0.7 for $k_{\perp max}=5$ GeV. The situation
is even worse in the case of the helix field where the expected number
of soft gluons per unit of rapidity is significantly larger.
We conclude that gluons softer than $k_{\perp 0}$ cannot be
implemented inside the Lund Model as individual gluonic excitations of
the string. 

We will end this appendix with an interpretation of the Lund
fragmentation model, which provides us with the possibility to relate
$k_{\perp 0}$ to the $b$-parameter in the model. The result in
Eq.(\ref{e:Lundprob}) (although derived semi-classically) can be
interpreted quantum-mechanically by a comparison to Fermi's Golden
Rule. It equals the final state phase space times the square of a
transition matrix element $|{\cal M}|^2 = \exp(-b A)$. There are two
such quantum-mechanical processes, Schwinger tunneling and the Wilson
loop integrals, which can be used in this connection (and they result
in very similar interpretations of the parameters).  For the Schwinger
tunneling case we note that if a constant ($\kappa$) force field is
spanned across the longitudinal region $X$ during the time $T$ with a
transverse size $A_{\perp}$ then the persistence probability of the
vacuum (i.e. the probability that the vacuum should not decay by the
production of new quanta) is~\cite{r:schwinger}
\begin{eqnarray}
\label{e:Schwingertunnel}
|{\cal M}|^2 = \exp(-\kappa^2XT A_{\perp} \Pi) \; \; .
\end{eqnarray}
Here the number $\Pi$ only depends upon the properties of the quanta
coupled to the field; for two massless spin $1/2$ flavours it is
$\Pi=1/12 \pi$.  Comparing the result in Eq.(\ref{e:Schwingertunnel})
to Eq.(\ref{e:Lundprob}) we find that the parameter~$b=A_{\perp}/24
\pi$ (taking into account that the Lund model area is counted in
lightcone units).
From Eq.(\ref{e:glueformation}) we obtain the minimum transverse size
of the field from which it then follows that the $b$-parameter in the
Lund model must be $b \geq \pi/24
\kappa \simeq 0.6$ GeV$^{-2}$. This is evidently just in accordance 
with the phenomenological findings in the Lund model for the parameter~$b$.
Further, considering the distribution in Eq.(\ref{e:pttunnel}) for the
transverse momentum of a produced $\mbox{q} \overline{\mbox{q}}$-pair
breaking the string we recognise the quantity $\ell_{min}^2$ in the
exponential fall-off.  We may conclude that there is a wave-function
for the Lund string in transverse space with just the right transverse
size to allow the ``ordinary'' transverse fluctuations in momenta.

\end{document}